# Temperature-driven Emergence of Negative Photoconductivity in Semimetal MoTe$_2$ Film Probed with Terahertz Spectroscopy

Jiaming Chen[1], Peng Suo[1], Wenjie Zhang[1], Hong Ma[2], Jibo Fu[1], Di Li[1], Xian Lin[1], Xiaona Yan[1], Zuanming Jin[3,6*], Weiming Liu,[4,6] Guo-Hong Ma[1,6*], Jianquan Yao[5]

[1]Department of Physics, Shanghai University, Shanghai 200444, China

[2]School of Physics and Electronics, Shandong Normal University, Jinan 250014, China

[3]Terahertz Technology Innovation Research Institute, Terahertz Spectrum and Imaging Technology Cooperative Innovation Center, Shanghai Key Lab of Modern Optical System, University of Shanghai for Science and Technology, 516 JunGong Road, Shanghai 200093, China

[4]School of Physical Science and Technology, ShanghaiTech University, Shanghai 201210, China

[5]Institute of Laser and Opto-electronics, College of Precision Instrument and Opto-electronics Engineering, Tianjin University, Tianjin 300072, China

[6]STU & SIOM Joint Laboratory for Superintense Lasers and the Applications, Shanghai 201210, China

*Corresponding authors:

physics_jzm@usst.edu.cn (Z. M. Jin), and ghma@staff.shu.dedu.cn (G. H. Ma),

ABSTRACT: Layered two-dimensional (2D) materials MoTe$_2$ have been paid special attention due to the rich optoelectronic properties with various phases. The nonequilibrium carrier dynamics as well as its temperature dependence in MoTe$_2$ are of prime importance, as it can shed light on understanding the anomalous optical response and potential applications in far infrared (IR) photodetection. Hereby, we employ time-resolved terahertz (THz) spectroscopy to study the temperature dependent nonequilibrium carrier dynamics in MoTe$_2$ films. After photoexcitation of 1.59 eV, the 1T'-phase MoTe$_2$ at high temperature behaves only THz positive photoconductivity (PPC) with relaxation time of less than 1 ps. In contrast, the T$_d$-phase MoTe$_2$ at low temperature shows an ultrafast THz PPC initially followed by emerging THz negative photoconductivity (NPC), and the THz NPC signal relaxes to the equilibrium state in hundreds of ps time scale. Small polaron formation induced by hot carrier has been proposed to be ascribed to the THz NPC in the polar semimetal MoTe$_2$ at low temperature. The polaron formation time after photoexcitation increases slightly with temperature, which is determined to be ~0.4 ps at 5 K and 0.5 ps at 100 K. Our experimental result demonstrates for the first time the dynamical formation of small polaron in MoTe$_2$ Weyl semimetal, this is fundamental importance on the understanding the temperature dependent electron-phonon coupling and quantum phase transition, as well as the designing the MoTe$_2$-based far IR photodetector.

## Introduction

Transition-metal dichalcogenides (TMDs) have aroused considerable interests due to their excellent optical performance, high thermoelectric properties, robust stability in air and promise for practical applications.[1-5] Among abundant TMDs family members, $MoTe_2$ stands out from others due to the presence of rich structural and electronic phases, resulting in the fascinating electrical and optical properties.[6-7] Conventionally, $MoTe_2$ crystallizes into three structural phases: a hexagonal 2H semiconducting phase, a centrosymmetric monoclinic 1T' semimetal phase, and an orthorhombic $T_d$ semimetal phase.[8-11] Among the semimetal phase of $MoTe_2$, the 1T' phase is a quantum spin Hall insulator in the few-layer form, while the $T_d$ phase is a type-II Weyl semimetal candidate due to the breaking of lattice inversion symmetry. The centrosymmetric 1T' $MoTe_2$ phase can be transformed into the noncentrosymmetric $T_d$ phase by cooling the compound down to the temperature of 240 K.[12-14] It has been reported that $T_d$-$MoTe_2$ exhibits extremely large magnetoresistance arising from the combination of the electron-hole compensation and a particular texture on the electron pockets.[15] The $T_d$-$MoTe_2$ is also reported to be superconducting with $T_C$=0.1 K,[16-17] and the superconductivity is believed to arise from the enhanced electron-phonon (*e-ph*) coupling at low temperature. Considering the small energy difference between 1T' and $T_d$ phases, Zhang *et al.* have successfully demonstrated the ultrafast laser pulse driven lattice symmetry switch in $MoTe_2$ via coherently exciting the interlayer shear phonon mode.[18] By using Raman spectroscopy, Paul *et al*. have observed clear signature of *e-ph* coupling in $MoTe_2$ when phase transition takes place between 1T' and $T_d$,[11] which indicates that the *e-ph* coupling can be significant difference for the two semimetal phases. As a matter of fact, the centrosymmetric 1T' phase of $MoTe_2$ shows non-polar character, while noncentrosymmetric $T_d$ phase is a polar metal in nature,[12-13, 19] thus, the different *e-ph* coupling is expectable for the two semimetal phases of $MoTe_2$, which may lead to large difference in optical response.

Although extensive studies have been carried out on $MoTe_2$ in the past a few years, seldom is done in the THz frequency. As a type-II Weyl semimetal,[20] $MoTe_2$ is expected to be good candidate for long-wave photodetection, especially for THz radiation.[21-22] On the other hand, THz spectroscopy has noninvasive and contact-free characteristic, which is a powerful tool for investigating the nonequilibrium carrier relaxation. Moreover, the THz spectroscopy also provides an alternative way for obtaining microscopic parameters such as carrier mobility and momentum scattering time of a material, thus the scattering mechanism that dominated the relaxation of nonequilibrium states, for instance, *e-ph* coupling,[23] trion,[24] band gap opening,[25] and interfacial trapping[4] *etc*. can be clarified and expounded.

In this article, we employ time-resolved THz spectroscopy to study the photocarrier dynamics of semimetal $MoTe_2$ phase with various temperatures. We show that the 1T' $MoTe_2$ at high temperature exhibits ultrafast THz PPC after photoexcitation at 780 nm, which shows identical photoresponse in semiconducting 2H $MoTe_2$. The subsequent relaxation completes within 1 ps, which is ascribed to electron cooling process via *e-ph* coupling. In contrast, after photoexcitation the $T_d$-phase $MoTe_2$ at low temperature shows ultrafast THz PPC initially followed by emerging THz NPC signal that lasts in hundreds of ps time scale. We proposed that the THz NPC arises from the formation of small polaron with photoexcitation in the polar semimetal $MoTe_2$ at low temperature. The polaron formation time is determined to be 0.4 ps at 5 K to 0.5 ps at 100 K, increasing slightly with temperature. And the dissociation time of the polaron increases with decreasing temperature, falling in hundreds of ps time scale.

## Experimental details

**Sample characterization.** The high quality $MoTe_2$ films with thickness of ~20 nm on sapphire substrate without intended doping were synthesized by chemical vapor deposition (CVD) method (provided by SixCarbon company, Shenzhen, China). Figures 1(a) and (b) display room-temperature centrosymmetric monoclinic 1T' structure and low-temperature noncentrosymmetric orthorhombic $T_d$ structures, respectively. The structures between the two phases are analogous, the only difference is that the $T_d$ phase $MoTe_2$ has a 4° tilting angle from the stacking direction, c-axis of crystal. The background carrier density at room temperature for 1T' phase is $3.7×10^{22}$ cm$^{-3}$ estimated from THz time-domain spectroscopy, and the detailed calculation of carrier density is presented in note 1 of Supplementary Information (SI). Figure 1(c) shows the Raman shift in the spectral region from 50 to 250 cm$^{-1}$ for the film at room temperature and 80 K. It is obviously seen that the room temperature modes centered around 127 cm$^{-1}$ ($P_A$) and 108 cm$^{-1}$ ($P_B$) undergo mode splitting at 80 K, in which the $P_A$ mode is split into 126 cm$^{-1}$ and 130 cm$^{-1}$, while the PB mode is split into 106 cm$^{-1}$ and 112 cm$^{-1}$, as illustrated in Figure 1(d) and Figure 1(e), respectively. The mode splitting is ascribed to the contribution of $T_d$-MoTe$_2$ phases at low temperature,[11,26] indicating phase transition occurs from high temperature centrosymmetric 1T' to low temperature noncentrosymmetrytic $T_d$ phases.

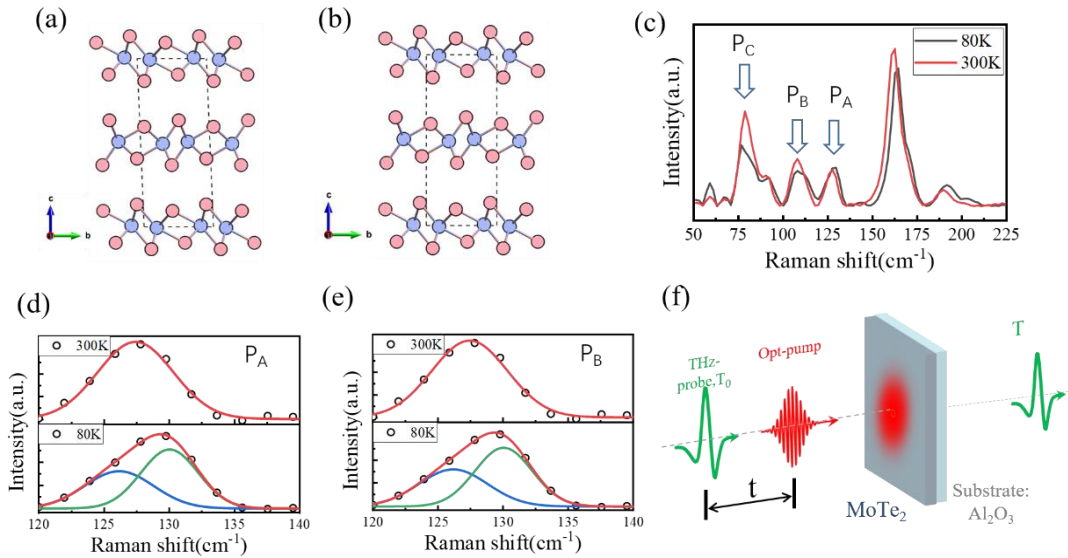

FIG 1. Characterization of MoTe$_2$ thin film. Molecular structure of 1T'-MoTe$_2$ (a) and $T_d$-MoTe$_2$ (b). Blue and pink spheres represent the Mo and Te atoms, respectively. (c) Raman spectra of MoTe$_2$ film at 80 K (black) and 300 K (red), the arrows mark the Raman shift of $P_A$, $P_B$ and $P_C$ peaks that are relevant with temperature induced phase transition as discussed in the text. (d) and (e) Enlargement of Raman spectra around 127 cm$^{-1}$ and 108 cm$^{-1}$ at 300 K (up) and 80 K (bottom) along with Gaussian fitting (solid line). (f) Schematic diagram of optical pump-THz probe spectroscopy.

**Transient THz dynamics measurement.** The time-resolved optical pump and THz probe (OPTP) experiments in transmission configuration were performed to explore the dynamics of photocarriers, which is illustrated in Figure 1(f).[27] The optical pulses are delivered from a Ti: sapphire amplifier with 120 femtoseconds (fs) duration at central wavelength of 780 nm (1.59 eV) and a repetition rate of 1 kHz. The THz emitter and detector are based on a pair of (110)-oriented ZnTe crystals. The optical pump and THz probe pulse are collinearly polarized with a spot size of 6.5 and 3.0 mm on the surface of sample, respectively. All measurements were conducted in dry nitrogen atmosphere to avoid the vapor absorption, and the samples were placed in a cryostat with temperature varying from 5 K to 300 K. Figure 1(f) shows schematically the experimental setup for optical pump and THz probe spectroscopy.

To unravel the photocarrier dynamics in THz frequency, both semiconducting and semimetal phases of $MoTe_2$ films with identical thickness of 20 nm are used for this study, and we mainly concentrated on the temperature dependent THz PC of semimetal $MoTe_2$ film in this study. The room temperature X-ray diffraction (XRD) pattern as well as the UV-visible absorption spectra for both 2H and 1T'-$MoTe_2$ films are presented in Figure S2 of note 2 in SI. As displayed in Figure S2(a), the room temperature XRD pattern exhibits pronounced (002) and (004) characteristic peaks along with two weak peaks indexed as (006) and (008) for both films, indicating good crystallinity of $MoTe_2$ films. It is also clearly seen from Figure S2(b) that the 1T' film is zero band gap indicating metallic phase in nature, while the bandgap of the 20 nm 2H $MoTe_2$ film is determined to be 0.75 eV. Previous reports indicate the monolayer a 2H $MoTe_2$ has a direct band gap structure with band gap of ~1.10 eV.[28] With increasing the film thickness, the 2H $MoTe_2$ undergoes an indirect-gap semiconductor with band gap of ~0.8-0.9 eV depending on the film thickness, substrate stress and fabrication method. [25,29] The 0.75 eV band gap measured in our 2H $MoTe_2$ film might arise from the film fabricated by CVD method as well as the sapphire substrate used.

## Experimental results

In order to study the photocarrier dynamics of $MoTe_2$ films, we have applied THz pulse after an optical pump at 780 nm and measured the photoinduced THz transmission change, which is defined as $\Delta T=T-T_0$ with respect to the delay time, $\Delta t$, between THz pulse and optical pulse, where the $T_0$ and $T$ denote the transmission signals of THz electric field peak value without and with photoexcitation, respectively. In a thin film approximation, the pump induced THz transmission change $\Delta T/T_0$, is proportional to the negative THz PC $-\Delta\sigma$, i.e. $\Delta T/T_0 \sim -\Delta\sigma$.

Figure 2 (a) shows the transient THz transmission of the semimetal MoTe$_2$ film at 5 K and room temperature under the same pump fluence of 0.36 mJ/cm$^2$. The transient THz response at room temperature shows sharp decrease in THz transmission after photoexcitation, the subsequent relaxation can be reproduced with a biexponential function with typical time constants of $\tau_1$=0.55 ps and $\tau_2$=5.6 ps. It should be noted that the amplitude ratio in $\tau_2$-component is less than 6% of that in $\tau_1$-component as shown in note 3 of SI, which indicates that the fast component plays a dominated role in the photocarrier relaxation for the semimetal MoTe$_2$ film. In contrast, the transient THz transmission of MoTe$_2$ at 5 K exhibits quite different behavior as that of room temperature one. It is clear that the transient THz PC at low temperature consists of three stages with distinct time scales: (1) a pump-induced sharp increase in THz PC (drop in THz amplitude transmission) with response time limited by the laser pulse duration; (2) the transformation from THz absorption to THz bleach that achieves maximum magnitude of bleach with time constant of a few ps; (3) the slow recovery process lasts hundreds of p*s* from the maximum bleach signal to equilibrium state. For comparison, we also present the transient THz response of 2H-MoTe$_2$ film under identical experimental condition. It is clear that no observable bleach of THz transmission occurs at both room and low temperatures for 2H MoTe$_2$ film, and the relaxation time can be well reproduced with a single exponential decay with fitting time constant of ~3.0 ps at 5 K and ~3.5 ps at 300 K. It should be noted that the relaxation time in 2H MoTe$_2$ film is close to the time constant of $\tau_2$-component measured in semimetal MoTe$_2$ film. We ascribed the $\tau_2$-component in semimetal MoTe$_2$ film to the contribution of 2H MoTe$_2$ phase, we have identified that the semimetal MoTe$_2$ film is slightly doped with tiny 2H phase MoTe$_2$ during the film growth with CVD method, which is revealed by optical image and Raman spectroscopy as illustrated in Figure S7 in note 7 of SI, and the detailed explanation will be presented in the discussion part later.

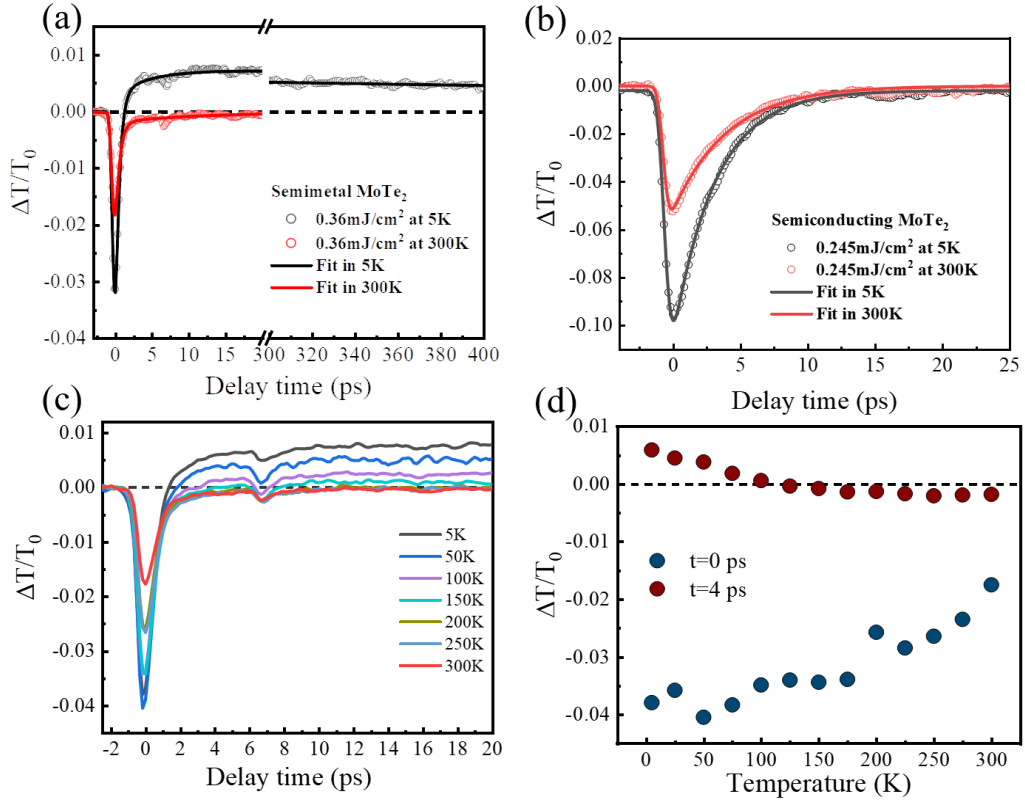

FIG 2 Impact of temperature on the transient THz transmission of MoTe$_2$ film. (a) Transient THz transmission of 1T'-MoTe$_2$ film at 300 K (red) and 5 K (black) with the same pump fluence of 0.36 mJ/cm$^2$. The solid lines are fitting curves with biexponenetial (red) and triexponential (black). (b) transient THz transmission of 2H-MoTe$_2$ film at 300 K (red) and 5 K (black). The solid lines are fitting curves with single exponential. (c) Transient THz transmission of semimetal MoTe$_2$ film under various temperatures under photoexcitation at 780 nm. (d) Magnitude of $\Delta T/T_0$ collected at time delay $\Delta t$=0 ps (black) and 4 ps (red), respectively. The dashed line denotes the baseline, i.e. $\Delta T/T_0$=0.

At first, we discuss the originality of the THz PPC after photoexcitation for MoTe$_2$ films. The THz PPC can arise from the photoinduced increase of photocarrier, and this is the case for the semiconducting MoTe$_2$ film shown in Figure 2(b). On the other hand, THz PPC can also arise from the carrier thermalization in semimetal, like the cases of semimetal Cd$_3$As$_2$ and PtTe$_2$.[23,29] As the metal-like property with high carrier concentration (~10$^{22}$ cm$^{-3}$) in our semimetal MoTe$_2$ film, the photoexcitation mainly results in the thermalization of electrons through electron-electron (*e-e*) scattering within several tens of fs, and the increase of carrier density (~$10^{19}$ cm$^{-3}$) is negligible. At elevated electron temperature caused by optical injection, the increase in Drude weight is much larger than that in scattering rate due to the constraint of linear dispersion band structure in a 3D Weyl semimetal, like MoTe$_2$. As a result, the free carriers' absorption of THz pulses is enhanced by hot electrons due to the intraband transition undergoes larger possible momentum and energy conservation spaces[23,30-31]. Therefore, the sharp enhancement of THz absorption in our semimetal MoTe$_2$ film originates from the production of hot electrons.

The most pronounced feature in Figure 2(a) is the appearance of THz NPC after photoexcitation, and this NPC only takes place in semimetal MoTe$_2$ at low temperature. It is noted that a first-order phase transition is involved by cooling bulk MoTe$_2$ down to 240 K and below, in which MoTe$_2$ is transformed from high temperature 1T' phase to low temperature T$_d$ phase.[10-11] To clarify the film undergoing a phase transition during the cooling process, we have measured the Raman spectra of the film at 300 K and 80 K, as shown in Figure 1(c), respectively. The mode splitting is seen for both P$_A$ and P$_B$ peaks at 80 K, as illustrated in Figure 1(d) and 1(e). Previous studies have confirmed that the Raman peak at these positions undergo splitting when the phase transition occurs.[11] The emerging THz NPC after photoexcitation is expected to arise from the low temperature T$_d$-MoTe$_2$ phase. It is clear that the THz PC is closely related to the photocarrier density (Δn) as well as the photoinduced mobility change (Δμ), i.e. (Δσ=Δneμ+neΔμ). Photoexcitation causing increase in carrier density is negligible comparing with the intrinsic carrier density in our semimetal film, moreover, the photoexcitation induced increase in carrier density can't lead to the THz NPC, thus the NPC can only root from the reduction of carrier mobility. To clarify this process, we have tested the transient THz transmission in the temperature range of 5-300 K, as shown in Figure 2(c). The THz NPC signal exhibits an obvious temperature dependence, more negative magnitude becomes with lower temperature. Figure 2(d) plots the transmitted THz amplitude with respect to temperature measured at delay time of Δt=0 ps (dark) and Δt=4 ps (red). The dark dots indicate that THz absorption (PPC) increases with decreasing temperature after photoexcitation. The magnitude of bleaching signal increases with decreasing temperature, and the bleaching signal is only observable when the temperature is below ~200 K, which manifests that the THz NPC could originate from the $T_d$-phase MoTe$_2$.

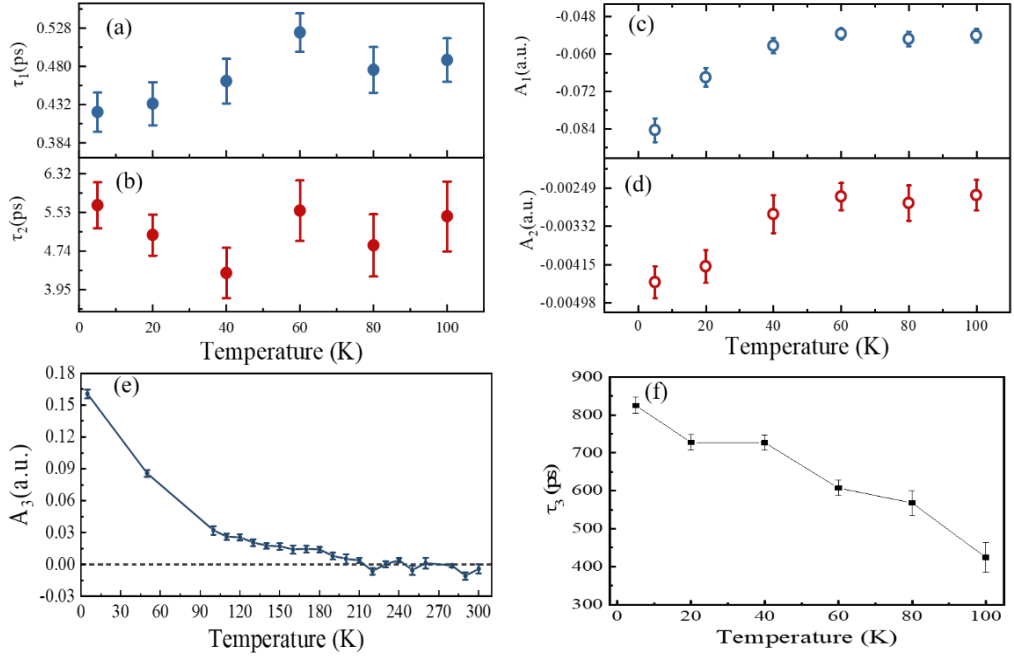

FIG 3. The fitting parameters for the temperature dependent transient THz transmission of semimetal MoTe$_2$. (a) and (b) The fitting decay time constants of $\tau_1$ and $\tau_2$ with respect to temperature under 780 nm excitation. (c) and (d) The fitting amplitude of $A_1$ and $A_2$ with respect to temperature. (e) The fitting amplitude of $A_3$ with respect to temperature, and (f) the fitting decay time constant of τ3 with respect to temperature.

To gain the recovery process of the bleaching signal at low temperature, we have measured the transient THz transmission in long scanning time window under various temperatures, which is presented in note 4 of Figure S4 in SI. Here, we use a triple-exponential decay convoluted with a Gaussian function to fit the experimental results measured at various temperatures and fluences,

$$\frac{\Delta T}{T_0}(t) = A_1 \cdot e^{\frac{\omega^2}{\tau_1^2} - \frac{t}{\tau_1}} \cdot erfc\left(\frac{\omega}{\tau_1} - \frac{t}{2\omega}\right) + A_2 \cdot e^{\frac{\omega^2}{\tau_2^2} - \frac{t}{\tau_2}} \cdot erfc\left(\frac{\omega}{\tau_2} - \frac{t}{2\omega}\right) + A_3 \cdot e^{\frac{\omega^2}{\tau_3^2} - \frac{t}{\tau_3}} \cdot erfc\left(\frac{\omega}{\tau_3} - \frac{t}{2\omega}\right) \quad (1)$$

in which $\tau_i$ (*i*=1, 2, 3) denotes for the time constant of relaxation process, $A_i$ (*i*=1, 2, 3)

stands for the corresponding amplitude, ω is the full width at half maximum (FWHM) of THz waveform, and *erfc(x)*=1-*erf(x)* is the complementary error function. The experimental data can well be reproduced with triple-exponential function, which are shown in solid lines in Figure S4 of note 4 in SI. The fitting parameters with respect to temperature are shown in Figure 3. After photoexcitation, the relaxation dynamics is composed of three processes when the temperature is below 200 K. When the temperature is higher than 200 K, the THz bleaching signal disappears, and the relaxation process can be well fitted with a biexponential function (i.e. $A_3$=0 in Eq. (1)). It is seen from Figure 3(a) and (b), the time constant of fast process increases from 0.42 ps at 5 K to 0.52 ps at 100 K, while the lifetime of slow component varies from 4.3 ps at 5 K to 5.6 ps at 100 K. The more fitting parameters at higher temperature are presented in note 5 of Figure S5 in SI. Figure 3(c) and (d) presents the corresponding amplitudes, $A_1$ and $A_2$, from which it is clearly seen that the magnitudes of both |$A_1$| and |$A_2$| decrease with increasing temperature, and obviously the ratio of $|A_2/A_1|$<6% that indicates the fast process plays dominated role for the positive THz PC relaxation. This process comes from the photocarrier cooling by transferring their excess energy to the lattice via *e-ph* interactions. The a-few-ps slow process with much smaller magnitude comes from the contribution of the tiny semiconducting impurity in the semimetal film. The originality of the semiconducting impurity is assigned as the presence of 2H phase MoTe$_2$ during the film growth by CVD method. Figure S6 in SI provides micro-optical image and Raman spectral evidences for the existence of tiny 2H phase in the semimetal film.

Figure 3(e) and (f) plots the fitting amplitude of $A_3$ as well as the corresponding lifetime τ$_3$ with respect to temperature. It is clear that the amplitude $A_3$ approaches zero at temperature around T$_{cr}$~ 210 K and above. This critical temperature, T$_{cr}$ is also close to the phase transition temperature we measured with temperature dependent THz transmission, which is displayed in Figure S7 in SI. The transition temperature of 210 K here is lower than 240~250 K reported in literatures,[11,14,18] the deviation of the observed T$_{cr}$ from other reports could be due to the much thinner MoTe$_2$ film (~20 nm) used in our study compared to the bulk material. But this does not hinder the observation of THz NPC at low temperature. In a word, when temperature is lower than ~200 K, a clear photoinduced THz NPC is observed, and the recovery time of THz NPC shows strongly temperature dependence, which changes from ~825 ps at 5 K to 424 ps at 100 K. The lifetime of $\tau_3$ can't be resolved clearly as the temperature is higher than 100 K due to the very small THz bleaching signal at higher temperature.

We further investigate the pump fluence dependence of the transient THz transmission at 5 K, as shown in Figure 4(a). The increase of pump fluence enhances the heating effect on the carriers, resulting in a greater THz absorption, the inset in Figure 4(a) shows the THz transmission, ΔT/T, at Δt=0 and 4 ps with respect to the pump fluence, respectively. Figures 4(b) and (c) show the fitting lifetimes and the corresponding amplitudes for the fast and slow components, respectively. Having shown in Figure 4(b), the lifetime of the fast component increases with pump flunece, while the lifetime of the slow component remains almost unchanged with increase of pump fluence. This further demonstrates that the fast component arises from electron cooling process via *e-ph* coupling, and the slow one is contributed by the photocarrier trapping by defect states in semiconducting phase of MoTe$_2$. Figure 4(c) shows the fitted fast component magnitude, $|A_1|$, with weight of ~85% increases linearly with the pump fluence, and the pump fluence dependent magnitude of slow component, $|A_2|$, with a weight of ~5% shows much smaller slope than the case of pump fluence dependent magnitude of $|A_1|$, which indicates the fast component plays a dominated role in the relaxation of positive THz PC .

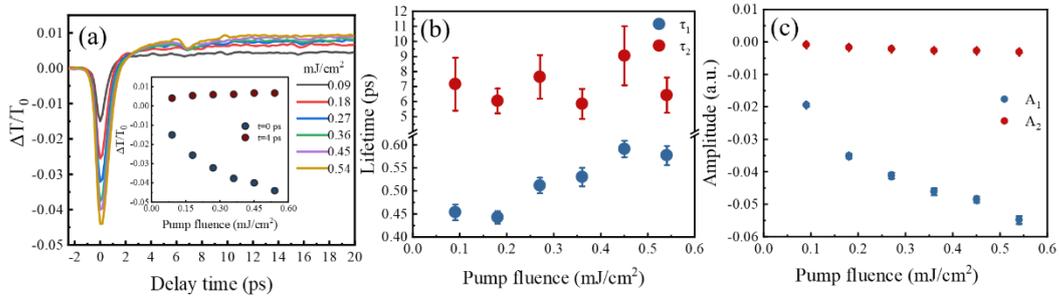

FIG4. Pump fluence dependence of transient THz transmission of MoTe$_2$ at 5 K. (a) Pump fluence dependence of transient THz transmission of semimetal MoTe$_2$ at 5 K under pumping wavelength of 780 nm. (b) The fitted time constants of $\tau_1$ (fast component) and $\tau_2$ (slow component) with respect to pump fluence. (c) The fitted amplitudes of $A_1$ and $A_2$ with respect to pump fluence.

Discussions

The observation of the NPC in THz frequency has been widely reported, such as graphene,[32-33] $Bi_2Se_3$,[34-35] $MoS_2$,[24] and carbon nanotube,[36] which were proposed to arise from the role of hot electrons, metallic surface states, formation of trions, respectively. Here we assigned the negative THz PC to the excitation of small polaron in the low temperature phase $MoTe_2$. In order to consolidate the assignment of the small polaron formation after photoexcitation, we discuss and rule out other possible origins of the negative THz PC. Firstly, photoexcitation leading to the elevated lattice temperature may be a possible contribution of the observed negative PC in $MoTe_2$ film. We rule out this possibility based on following two facts: one is that the THz NPC is only observed at low temperature phase, *i.e.* $T_d$-phase, which disappears at the high-temperature 1T'-phase. Considering both $T_d$- and 1T'-phase $MoTe_2$ are semimetal with similar THz conductivity, it is reasonable to assume that heated lattice in $T_d$- and 1T'-phase $MoTe_2$ should have identical THz response after photoexcitation, therefore it is impossible that the low temperature phase $MoTe_2$ shows a pronounced NPC, while the high temperature phase $MoTe_2$ does not. The other is that according to two-temperature model, the *e-e* thermalization is much faster than *e-ph* thermalization after photoexcitation, and the electrons with the elevated temperature transfer the excess energy to the lattice via *e-ph* coupling until the two subsystems reach a balanced temperature. Then, the long-lived bleaching signal could arise from the heat dissipation to substrate via phonon-phonon coupling. If it is this case, the heat dissipation process to the substrate is expected to become slower with increasing temperature, this is because the temperature gradient between the heated film and substrate becomes smaller with increasing temperature. However, it is noted from Figure 3(f) that the recovery time of the bleaching signal shows decrease with rising temperature. Based on the discussions above, we can rule out the hot carrier contribution to the observed THz NPC signal with certainty. Another possibility is that the observed NPC in $MoTe_2$ arises from the formation of trions after photoexcitation as such case in monolayer $MoS_2$.[24] Considering the metallic nature of the film with high carrier concentration, the formation of the exciton with photoexcitation can be totally screened by the background free carriers, thus this possibility can be safely ruled out. In addition, no NPC observed in $MoTe_2$ at room further excludes the conjecture about the photoinduced formation of trions. Thirdly, the impurity in the $MoTe_2$ film may also lead to the THz NPC after photoexcitation as observed in some semiconductors.[37-38] It is noted that the semimetal $MoTe_2$ thin film grown by CVD method could include a few semiconducting (2H) phase, which may play a dominated role in the THz NPC after photoexcitation. While, the contribution of semiconducting phase to the NPC can be totally ruled out according to the experimental results shown in Figure 2(b), in which 2H phase $MoTe_2$ film only behave THz PPC at both high- and low-temperature. In addition, the relaxation time (*i.e.* hundreds of ps) of the negative THz PC observed in Figure 2(a) is 4~6 orders of magnitude faster than the recovering time of the NPC reported in semiconductors.[37-38] Fourthly, one may argue that 1T' and $T_d$ phases may co-exist at certain temperature, 1T'/$T_d$ interfacial charge transfer (CT) may occur from high conductivity phase to the low one with photoexcitation, as a result, the THz NPC may arise from the CT instead of formation of polaron. The CT contribution to the THz NPC can be ruled out by considering temperature dependent THz NPC signal: It is known that the phase transition occurs at critical temperature of $T_{cr}$~200 K for our semimetal $MoTe_2$ film, it is reasonable to assume that the quantity of 1T'/$T_d$ interface decreases with decreasing temperature when temperature is lower than $T_{cr}$. Therefore, more interfacial CT is anticipated to take place at higher temperature than that at 5 K under identical photoexcitation. As a result, the THz NPC signal at higher temperature (for example, 100 K) is then expected to be more pronounced than that at 5 K. Obviously, our experimental results show that lower temperature has larger NPC signal, which contradicts with the CT-based assumption. Last but not least, we noted that optical induced THz radiation under normal incident from our samples is negligible so that the observed negative THz PC signal can also rule out the contribution from the photo-induced THz radiation. Based on the analyses above, we interpret the unexpected NPC as the photo-excitation of small polaron in MoTe$_2$ at low temperature, in which the formation of small polaron leads to the significant reduction of carrier mobility due to the phonon "dressing" of carrier.[30]

The formation of polaron is triggered by hot carriers injected with photoexcitation,[30] the interaction between hot carriers and cold lattice via *e-ph* coupling leads to the distortion of lattice, as a result, the electrons trapped by distorted lattice move with much large effective mass, and therefore with much lower conductivity, which leads to the reduction of THz absorption and the increase of THz transmission. It is clear that formation of polaron is strongly relevant to the *e-ph* coupling of a material. The observation of superconducting property of T$_d$ MoTe$_2$ at T$_C$=0.1 K implies large *e-ph* coupling at low temperature,[17] and this pronounced *e-ph* coupling at low temperature phase is also evidenced by the anomalous behavior of P$_1$(79 cm$^{-1}$) peak in Raman studies,[11] which is in accordance with our temperature dependent Raman spectra of P$_C$ mode (see Figure S8 in SI). Moreover, it is noted that 1T'-phase MoTe$_2$ at high temperature has inversion symmetry, which is non-polar semimetal, as the temperature cooling down, MoTe$_2$ undergoes first order phase transition from non-polar 1T' phase to polar T$_d$ phase,[19] the emerging polarity in T$_d$ phase MoTe$_2$ tends to have larger *e-ph* coupling coefficient than that of the non-polar 1T' phase counterpart, so that it is more easily to form small polaron for the T$_d$ phase MoTe$_2$ with photoexcitation. In Addition, the pump fluence dependent THz NPC shown in Figure 4 does support the conjecture about the small polaron formation after photoexcitation. Higher pump fluence leads to higher electron temperature produced in the T$_d$ MoTe$_2$ film, therefore more phonon modes are excited during the hot electron cooling process via *e-ph* coupling due to the fact that the excited phonon modes are proportional to the temperature difference between the hot electron and the cold lattice. As a result, more electrons are "trapped" by phonons, which results in an enhanced negative THz PC signal under higher pump fluence. Recent experimental and theoretical studies reveal ultrafast laser excitation can enhance the *e-ph* coupling of a material.[39-43] By measuring the Raman spectrum of single layer graphene with ultrafast laser pulse, Ferrante *et al* demonstrated that the linewidth of both G and 2D peaks increase with pump fluence due to the light-enhanced *e-ph* coupling originating from the phonon-assisted intraband transitions.[39] By selectively exciting the infrared phonon mode of bilayer graphene, Pomarico *et al.* observed a clear signature of light-enhanced e-ph coupling.[41] A theoretical study by Kennes *et al* suggesting that transient electronic attraction arises from nonlinear *e-ph* coupling with specific phonon mode excitation.[42] Sentef suggested that the *e-ph* coupling in graphene can be strengthened by light illumination due to the presence of distorted lattice and phonon nonlinearity.[43] Utilizing various theoretical approaches, a common trend of increasing *e-ph* coupling with increasing electron temperature has been reached for a wide of metals, which indicates that the elevated electron temperature by ultrafast laser excitation can result in the enhancement of *e-ph* coupling in some metals.[44-46] In short, the photoinduced formation of small polaron in $T_d$ phase $MoTe_2$ is depicted as follows: upon photoexcitation, the distribution of nonequilibrium carriers is broadened at a larger density of states, stronger hot carrier-phonon interaction is in favor of lattice deformation, which leads to formation of metastable polaron in $T_d$ phase $MoTe_2$. The long-lived negative photoconductivity corresponds the dissociation process of the small polaron.

**In summary**, we have utilized ultrafast optical pump and THz probe spectroscopy with various pump fluences and temperatures to investigate the relaxation dynamics of photoexcited carriers in semimetal $MoTe_2$ film. We ensure that the initial fast relaxation of hot electrons is due to the *e-ph* coupling, transferring energy to the lattice. With the temperature cooling down, semimetal $MoTe_2$ film undergoes a phase transition from 1T' to $T_d$, and the significant feature of the transient THz transmission at $T_d$ $MoTe_2$ is the enhanced THz absorption followed by a THz photobleaching signal with lifetime of hundreds of ps. We proposed that the THz NPC at low temperature arises from the small polaron formation after photoexcitation via strong *e-ph* coupling in the polar semimetal of $T_d$ $MoTe_2$ film. The polaron formation time is fitted to be less than 1 ps that is dominated by the *e-ph* coupling, and the very slow recovery of bleaching signal with time constant of hundreds of ps comes from the dissociation of polaron that is strongly related to temperature. Our investigations open the door to understand the photocarriers' dynamics in type-II Weyl semimetal and offering insights into the optoelectronic device applications.

**Acknowledgements**

This work was supported by the National Natural Science Foundation of China (NSFC, Nos. 92150101, 61735010, and 61975110).